\documentclass[journal=jpcafh,manuscript=article]{achemso}
\usepackage{amsmath}
\usepackage{amssymb}
\usepackage[version=4]{mhchem}
\usepackage{xcolor}
\usepackage{hyperref}
\usepackage{mathtools,xparse}
\usepackage{placeins}
\usepackage{todonotes}
\setkeys{acs}{maxauthors = 0}

\newcommand{\ket}[1]{ \left| #1 \right \rangle}
\newcommand{\bra}[1]{ \left\langle #1 \right |}

\makeatletter
\DeclareFontFamily{OMX}{MnSymbolE}{}
\DeclareSymbolFont{MnLargeSymbols}{OMX}{MnSymbolE}{m}{n}
\SetSymbolFont{MnLargeSymbols}{bold}{OMX}{MnSymbolE}{b}{n}
\DeclareFontShape{OMX}{MnSymbolE}{m}{n}{
    <-6>  MnSymbolE5
   <6-7>  MnSymbolE6
   <7-8>  MnSymbolE7
   <8-9>  MnSymbolE8
   <9-10> MnSymbolE9
  <10-12> MnSymbolE10
  <12->   MnSymbolE12
}{}
\DeclareFontShape{OMX}{MnSymbolE}{b}{n}{
    <-6>  MnSymbolE-Bold5
   <6-7>  MnSymbolE-Bold6
   <7-8>  MnSymbolE-Bold7
   <8-9>  MnSymbolE-Bold8
   <9-10> MnSymbolE-Bold9
  <10-12> MnSymbolE-Bold10
  <12->   MnSymbolE-Bold12
}{}

\let\llangle\@undefined
\let\rrangle\@undefined
\DeclareMathDelimiter{\llangle}{\mathopen}%
                     {MnLargeSymbols}{'164}{MnLargeSymbols}{'164}
\DeclareMathDelimiter{\rrangle}{\mathclose}%
                     {MnLargeSymbols}{'171}{MnLargeSymbols}{'171}
\makeatother

\author{Peter Reinholdt}
\affiliation[SDU]{Department of Physics, Chemistry and Pharmacy, University of Southern Denmark, Campusvej~55, DK--5230 Odense M, Denmark}
\email{reinholdt@sdu.dk}

\author{Erik Rosendahl Kjellgren}
\affiliation[SDU]{Department of Physics, Chemistry and Pharmacy, University of Southern Denmark, Campusvej~55, DK--5230 Odense M, Denmark}

\author{Jacob Kongsted}
\affiliation[SDU]{Department of Physics, Chemistry and Pharmacy, University of Southern Denmark, Campusvej~55, DK--5230 Odense M, Denmark}

\title{Perturbatively Corrected Linear Response Selected Configuration Interaction}

\begin{document}

\begin{tocentry}
\includegraphics[width=8.25cm]{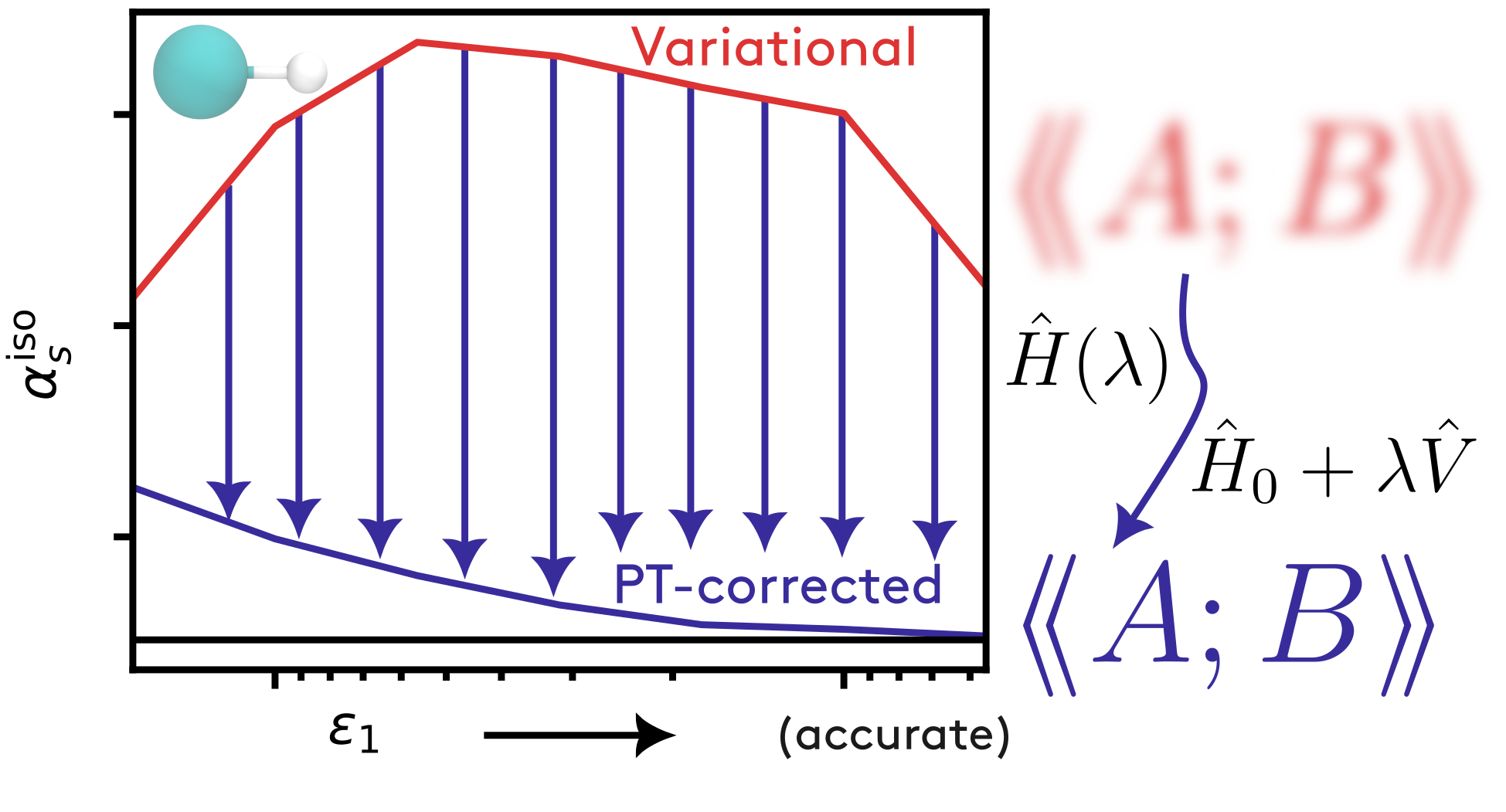}
\end{tocentry}

\begin{abstract}
Selected configuration interaction (SCI) methods have emerged as powerful, lower-cost alternatives to full configuration interaction (FCI) for ground- and excited-state energies. Still, calculating molecular response properties with SCI remains a significant challenge. 
In this work, we introduce perturbative corrections to the linear response selected configuration interaction (LR-SCI) framework, using an order-by-order Epstein-Nesbet perturbation expansion through second order.
We demonstrate that in this theoretical framework, the finite-order perturbative treatment preserves the pole structure of the parent variational LR-SCI theory, which means that although the method can be useful for static properties, it is not suitable for frequency-dependent molecular response properties.
Numerical benchmarks targeting the static polarizabilities of water, ethene, boron hydride, and hydrogen chloride demonstrate systematic convergence toward the FCI limit for both ground and excited electronic states. While first-order corrections yield marginal improvements, the inclusion of second-order corrections substantially enhances accuracy over underlying variational treatments and diminishes oscillatory convergence behavior present in the parent variational LR-SCI method. Combined with extrapolation techniques, LR-SCI-PT achieves excellent agreement with high-level coupled-cluster references, establishing a powerful route toward near-FCI quality molecular properties for systems otherwise inaccessible to exact FCI treatments.  
\end{abstract}

\section{Introduction}

One of the central challenges within molecular electronic structure theory is the development of methods capable of achieving results with high accuracy\cite{eriksen2020shape}.
The full configuration interaction (FCI) method gives the exact solution to the electronic Schr\"odinger equation within a given one-electron basis. However, the combinatorial growth of the determinant space with the number of electrons and orbitals means that exact FCI calculations are feasible only for relatively small molecular systems. 
Although large-scale exact FCI calculations have been reported with trillions of determinants\cite{gao2024distributed}, the steep computational scaling of the FCI method means that the actual size of systems accessible to the method remains rather small.

In many cases, exact FCI is not strictly required, and one may settle for much more affordable methods which provide near-FCI quality (e.g., within 1$\mu$H for energies), but at a significantly reduced cost\cite{loos2024go,craciunescu2025selected}. Many theoretical approaches aim at achieving this goal, with notable examples being the methods based on the density matrix renormalization group (DMRG)\cite{chan2011density}, many-body expanded FCI (MBE-FCI)\cite{eriksen2017virtual,eriksen2018many,eriksen2019many}, and selected configuration interaction (SCI)\cite{gershgorn1968application,huron1973iterative,caballol1992direct,tubman2016deterministic,holmes2016heat,liu2016ici}, among others\cite{motta2018ab,xu2018full,li2022downfolded}.

Ground-state energies are often the focus of such electronic structure methods, but many chemically relevant observables are related to molecular response properties rather than just the total energy\cite{helgaker2012recent}. 
Within SCI, some molecular properties are relatively easy to obtain, including excitation energies, which can be obtained from energy differences of low-lying eigenstates\cite{loos2018mountaineering}. 
SCI has been extensively used for benchmark calculations of excitation energies, for example, with the QUEST database \cite{loos2020quest,veril2021questdb,loos2025quest}, where SCI calculations combined with second-order perturbative corrections (SCI+PT2) and extrapolation techniques\cite{burton2024rationale} provide highly accurate estimates of FCI energies.
Expectation values and transition moments between different eigenstates, including dipole moments, oscillator strengths, and hyperfine coupling constants\cite{angeli1998multireference,angeli2001multireference,damour2022ground}, are also relatively straightforward to obtain.
However, treating molecular properties beyond expectation values or transition moments has yet to be explored in detail for SCI-type methods.

Molecular response properties describe how a molecular system responds to external perturbations, and include properties such as polarizabilities, magnetizabilities, NMR shielding tensors, and spin–spin coupling constants that cannot easily be obtained from ground-state energetics alone.
Although there have been some developments of approximate FCI methods capable of time-dependent or linear-response treatments\cite{schriber2019time,shee2025real,dorando2009analytic,nakatani2014linear,booth2012communication,blunt2015krylov,samanta2018response}, the development of response-theoretical frameworks for modern near-FCI methods remains comparatively limited. 

In a recent work\cite{reinholdt2025linear}, we presented an implementation of linear response for SCI wave functions (LR-SCI), capable of obtaining properties such as polarizabilities (including frequency-dependent and damped polarizabilities) or the response functions related to the calculation of NMR spin-spin coupling constants.
Within the original LR-SCI framework, we were able to obtain molecular response properties with an accuracy approaching the FCI limit, but at a significantly reduced cost.
Nonetheless, the original approach was still limited to relatively small molecular systems.

An analogous problem arises in conventional SCI calculations\cite{holmes2017excited,loos2018mountaineering} focused on obtaining ground- and excited-state energies, where a purely variational treatment also severely limits the method's size and scope. 
One of the most widely used approaches to address this limitation is to augment the purely variational treatment with a perturbative correction (i.e., SCI+PT) that accounts for the missing correlation energy from the determinants outside the variational space. 
Additionally, it has been recognized that convergence towards the FCI limit can be further accelerated with extrapolation techniques\cite{holmes2017excited,burton2024rationale}, which allow highly accurate estimates of the FCI energy from modestly sized variational expansions\cite{loos2018mountaineering}.
Motivated by the success of these ideas for ground-state energies, it is natural to consider whether similar perturbative corrections can be developed for molecular response properties within the LR-SCI framework.

\section{Theory}
\subsection{Selected Configuration Interaction}
In configuration interaction (CI), the electronic wave function $\ket{\Psi_0}$ is expressed as a linear combination of Slater determinants ${\ket{\Phi_I}}$ as
\begin{equation}
\ket{\Psi_0} = \sum_I c_I \ket{\Phi_I}, \label{eq:ci_wfn}
\end{equation}
where the coefficients $c_I$ are obtained by solving the time-independent electronic Schr\"odinger equation,
\begin{equation}
\hat{H} \ket{\Psi_0} = E_0 \ket{\Psi_0}.
\end{equation}
As written, Eq. \eqref{eq:ci_wfn} is so far slightly vague about which determinants are included in the summation. 
If all determinants are included, one arrives at the FCI approximation, which provides the exact solution to the electronic Schr\"odinger equation within a given basis. Although exactness is a desirable feature, the dimension of the FCI space grows combinatorially with the number of electrons and orbitals, which makes exact FCI applicable only to small molecular systems. 
One may alternatively turn to truncated CI hierarchies (e.g., CISD, CISDT), which reduce the computational cost by restricting the excitation level relative to a reference determinant.
However, truncated CI methods are not size-consistent\cite{bartlett1977determination} and often include many determinants with negligible coefficients\cite{ivanic2001identification}.

In SCI methods, one attempts to exploit the sparsity present in the FCI wave function by only including determinants that make a significant contribution to the wave function.
In practice, the selection of which determinants to include in the CI expansion is done iteratively: starting from some initial determinant set (usually the Hartree-Fock determinant), one solves the CI problem in the current variational space and uses the resulting wave function to identify and augment the set with determinants that are expected to contribute significantly to the wave function.
This process is repeated until some user-defined convergence threshold has been met. 
Various determinant selection strategies have been proposed \cite{huron1973iterative,tubman2016deterministic,holmes2016heat,liu2016ici}, but in this work, we shall focus on a version of Heat-Bath Configuration Interaction (HCI) \cite{holmes2016heat}. 
In HCI, a candidate determinant $\ket{\Phi_I}$ is added to the variational space if it satisfies
\begin{equation}
\left| H_{IJ} c_J \right| > \varepsilon_0, \label{eq:hci_criterion}
\end{equation}
for at least one determinant $\ket{\Phi_J}$ already present in the expansion. Here, $\varepsilon_0$ is a threshold that controls the size of the selected space, and  $H_{IJ}=\bra{\Phi_I}\hat{H}\ket{\Phi_J}$ is a Hamiltonian matrix element between determinants $\ket{\Phi_I}$ and $\ket{\Phi_J}$.
The HCI procedure is iterative. At each step, the Hamiltonian is diagonalized in the current determinant space to obtain the current set of CI coefficients. Next, new determinants are added according to the selection criterion, Eq. \eqref{eq:hci_criterion}. This process is repeated until an insignificant amount (e.g., 1\%) of additional determinants satisfy the criterion.

\subsection{Linear Response Theory}
Following the formalism of \citeauthor{koch1991analytical}\cite{koch1991analytical}, linear response properties for a CI wave function are determined by solving the linear response equations for the response vectors $\mathbf{X}_B(\pm \omega)$ 
\begin{equation}
\left( \mathbf{H} - (E_0 \pm \omega)\mathbf{I} \right) \mathbf{X}_B(\pm \omega) = \tilde{\mathbf{B}}. \label{eq:lr_equation}
\end{equation}
Here, $E_0$ is the ground-state energy, $\omega$ is the perturbation frequency, while the right-hand side $\tilde{\mathbf{B}}$ is the so-called property gradient (or property vector), which is obtained by multiplication of a one-electron operator $\hat{B}$, followed by projection of the reference state
\begin{equation}
\tilde{\mathbf{B}} = (\mathbf{I} - \mathbf{cc}^T)\mathbf{B}, \quad \text{with} \quad B_j = \langle \Phi_j | \hat{B} | \Psi_0 \rangle. \label{eq:property_vector}
\end{equation}
Here, we note that $\tilde{\mathbf{B}}$ is the property vector associated with the one-electron operator labeled $\hat{B}$, and that we may similarly encounter $\tilde{\mathbf{A}}$ as the property vector associated with the one-electron operator labeled $\hat{A}$.
The solution of the linear response equations typically uses iterative methods, e.g., Davidson-type schemes \cite{DAVIDSON197587,olsen1988solution}, which is also the approach we take in the present work.

After solving the linear response equations, the resulting response vectors can be used to evaluate linear response functions.
If we let $\hat{A}$ and $\hat{B}$ be components of the electric dipole operator, this allows the evaluation of the dipole-dipole polarizability  $\alpha_{AB}(\omega)$ as
\begin{equation}\alpha_{AB}(\omega) = -\llangle \hat{A}; \hat{B} \rrangle_\omega = \tilde{\mathbf{A}}^\dagger \left( \mathbf{X}_B(\omega) + \mathbf{X}_B(-\omega) \right) = \tilde{\mathbf{A}}^\dagger \mathbf{X}_{B,\pm\omega}, \label{eq:response_function}
\end{equation}
where we have introduced the shorthand notation $\mathbf{X}_{B,\pm\omega} = \mathbf{X}_B(\omega) + \mathbf{X}_B(-\omega)$.

As mentioned in Ref. \citenum{reinholdt2025linear}, the CI-based linear response equations described above are equally valid for an SCI treatment, which means that the practical implementation of LR-SCI is, in principle, straightforward.
However, an important practical issue concerning the determinant selection remains.
In conventional SCI, determinants are included in the CI expansion based on their expected importance in improving the ground-state energy using, for example, the HCI criterion in Eq.~\eqref{eq:hci_criterion}.
When targeting linear response properties, this set of determinants is not always appropriate, since determinants that are important for expanding property- and response vectors are not always contained within the ground-state (GS) set.
In Ref.~\citenum{reinholdt2025linear}, we addressed this problem by introducing two additional determinant addition criteria.
The first one (denoted +V) is aimed at achieving a good description of the property vectors, and adds determinants to the CI expansion if a candidate determinant $\ket{\Phi_I}$ satisfies 
\begin{equation}
\left|B_{IJ} c_J \right| > \varepsilon_V \label{eq:hci_prop_criterion}.
\end{equation}
Here, $B_{IJ}=\bra{\Phi_I}\hat{B}\ket{\Phi_J}$ is a matrix element of the one-electron operator $\hat{B}$ and $c_J$ is a CI wave function coefficient.
This selection criterion is an analogue to Eq.~\eqref{eq:hci_criterion}, but aims to ensure an accurate evaluation of one-electron matrix-vector products, which are the basis of the property vectors.
The second determinant criterion (denoted +X) aims to ensure a good description of the response vectors, and adds determinants to the CI expansion if it satisfies
\begin{equation}
\left|H_{IJ} X_J \right| > \varepsilon_X.\label{eq:hci_rsp_criterion}
\end{equation}
Here, $H_{IJ}$ is a Hamiltonian matrix element and $X_J$ is a component of the response vector. 
This criterion is thus analogous to the standard HCI criterion, but uses the coefficients from the response vector in place of the CI coefficients. This criterion is motivated by considering the linear response equation (Eq.~\eqref{eq:lr_equation}), where it ensures that matrix-vector products between the Hamiltonian and the response vector are accurately represented.
In practice, this second selection scheme is applied iteratively: following some initial solution to the linear response equations, additional determinants are added according to Eq.~\eqref{eq:hci_rsp_criterion}, and the linear response equations are solved again in the expanded space. This process is repeated until an insignificant number of determinants are added in the expansion step.
It is important to note that these determinant addition steps are always \emph{additive}, i.e., they are only used to expand the determinant space, not to retrospectively remove unimportant determinants.
In Ref.~\citenum{reinholdt2025linear}, we found that the combined use of both these selection criteria (denoted GS+V+X) was important for an accurate description of linear response properties with SCI. Accordingly, we shall use this combination throughout, and have applied a common threshold $\varepsilon_1 \equiv \varepsilon_0 = \varepsilon_V = \varepsilon_X$.

\subsection{Perturbation Corrections for the Energy and Wave Function}
The variational SCI wave function captures the dominant contributions to the electronic state, 
but there may still be a large number of neglected external determinants that give rise to non-negligible corrections. To account for these residual contributions in a computationally efficient manner, we introduce a perturbative treatment of the external space.

For this purpose, we introduce a partitioning of the Hamiltonian as
\begin{equation}
\hat{H}(\lambda) =  \hat{H}_0 +  \lambda \hat{V}. \label{eq:Hlambda}
\end{equation}
In the present work, we adopt an Epstein--Nesbet partitioning, in which the zeroth-order Hamiltonian $\hat{H}_0$ is defined to contain the full Hamiltonian within the variational space, while it is diagonal in the external space, with diagonal elements given by the corresponding Hamiltonian matrix elements. With this choice, the SCI wave function is an eigenfunction of $\hat{H}_0$ within the variational space. The perturbation operator $\hat{V} = \hat{H} - \hat{H}_0$ contains the remaining couplings, i.e., the off-diagonal matrix elements that connect the variational and external spaces, as well as the off-diagonal couplings within the external space.
Using an Epstein-Nesbet partitioning is a common choice for obtaining perturbative corrections to SCI wave functions\cite{holmes2016heat,garniron2019quantum}, though versions based on Møller-Plesset perturbation theory have also been explored\cite{angeli2002multireference}.

Having introduced the $\lambda$-dependent Hamiltonian in Eq.~\eqref{eq:Hlambda}, we proceed by inserting it into the Schr\"odinger equation
\begin{equation}
(\hat{H}_0 + \lambda \hat{V}) \ket{\Psi_0(\lambda)} = E_0(\lambda) \ket{\Psi_0(\lambda)}.
\end{equation}
Next, the energy and wave function are expanded as power series in $\lambda$
\begin{eqnarray}
    (\hat{H}_0 + \lambda \hat{V})  \left(\left| \Psi_0^{(0)} \right> + \lambda \left| \Psi_0^{(1)} \right> + \lambda^2 \left| \Psi_0^{(2)} \right> + ...\right) &=& \nonumber \\ ( E_0^{(0)} + \lambda E_0^{(1)} + \lambda^2 E_0^{(2)} + ...) \left(\left| \Psi_0^{(0)} \right> + \lambda \left| \Psi_0^{(1)} \right> + \lambda^2 \left| \Psi_0^{(2)} \right> + ...\right) && \label{eq:schrod_lambda}
\end{eqnarray}
The $n$'th order correction to the energy, $E^{(n)}$ can be obtained as
\begin{eqnarray}
    E_0^{(0)} &=& \left<\Psi_0^{(0)} \left|\hat{H}_0\right|\Psi_0^{(0)} \right> \\
    E_0^{(1)} &=& \left<\Psi_0^{(0)} \left|\hat{V}\right|\Psi_0^{(0)} \right>\\
    E_0^{(2)} &=& \left<\Psi_0^{(0)} \left|\hat{V}\right|\Psi_0^{(1)} \right>     
\end{eqnarray}
Here, we note that the first-order energy vanishes ($E_0^{(1)}=0$) due to the form of the Epstein--Nesbet partitioning since $\hat{V}\left|\Psi_0^{(0)}\right>$ is orthogonal to $\left|\Psi_0^{(0)}\right>$.
We also note that knowledge of the first-order coefficients is sufficient to evaluate the energy through second order.
    
Collecting orders in $\lambda$ in Eq. \eqref{eq:schrod_lambda}, we thus have the following linear equations for the wave function coefficients, expressed in the determinant basis
\begin{eqnarray}
     \mathbf{H}_0 \mathbf{c}^{(0)} &=& E_0^{(0)} \mathbf{c}^{(0)} \\
\mathbf{H}_0 \mathbf{c}^{(1)} + \mathbf{V}\mathbf{c}^{(0)}  &=&  E_0^{(0)} \mathbf{c}^{(1)} \\
 \mathbf{H}_0 \mathbf{c}^{(2)} + \mathbf{V}\mathbf{c}^{(1)} &=&  E_0^{(2)} \mathbf{c}^{(0)} + E_0^{(0)} \mathbf{c}^{(2)}
\end{eqnarray}
The zeroth-order wave function coefficients are identical to those obtained from the variational SCI calculation.
The first-order wave function coefficients are non-zero only in the external space and can be expressed algebraically as
\begin{equation}
    c_a^{(1)} = \frac{\langle \Phi_a | \hat{V} | \Psi_0^{(0)} \rangle}{E_0^{(0)} - H_{aa}},
\end{equation}
The second-order coefficients in the \emph{external} sector can similarly be formulated as
\begin{equation}
    c_a^{(2)} = \frac{\langle \Phi_a | \hat{V} | \Psi_0^{(1)} \rangle}{E_0^{(0)} - H_{aa}},
\end{equation}
while the second-order coefficients within the internal sector are obtained as the solution of the linear equation
\begin{equation}
    \left( \mathbf{H_0} - E_0^{(0)}\right) \mathbf{c}^{(2)} = E_0^{(2)} \mathbf{c}^{(0)} - \mathbf{V}\mathbf{c}^{(1)}. \label{eq:c2}
\end{equation}
In practice, the solution of Eq.~\eqref{eq:c2} is done using the same linear response solver routines used to solve the reference linear response equations.
At this point, we note that from the variational condition on the zeroth-order coefficients $(\mathbf{H}_0 - E_0^{(0)})\mathbf{c}^{(0)} = 0$, that if $\mathbf{c}^{(2)}$ is a solution of Eq. \eqref{eq:c2}, then $\mathbf{c}^{(2)} + \alpha \mathbf{c}^{(0)}$ will also be a solution. We fix this freedom by adopting intermediate normalization, i.e., requiring
\begin{equation}
    \left<\Psi_0^{(n)}|\Psi_0^{(0)}\right> = \delta_{n0}.
\end{equation}

\subsection{Perturbation Corrections for Linear Response Functions}
Having discussed the perturbation expansion of the wave function and ground-state energies, we next turn to consider the corresponding expansions of the linear response functions. 
In orders of $\lambda$, we have
\begin{equation}
    \llangle A ; B \rrangle _\omega(\lambda) =  \llangle A ; B \rrangle _\omega ^{(0)} + \lambda \llangle A ; B \rrangle _\omega ^{(1)} + \lambda^2 \llangle A ; B \rrangle _\omega ^{(2)} + ... 
\end{equation}
The property gradient can be expanded in orders of $\lambda$ as 
\begin{equation}
    \tilde{\mathbf{B}}(\lambda) = \tilde{\mathbf{B}}^{(0)} + \lambda \tilde{\mathbf{B}}^{(1)} + \lambda^2 \tilde{\mathbf{B}}^{(2)} + ...
\end{equation}
while the response vector can be expanded as 
\begin{equation}
    \mathbf{X}_{B,\pm\omega}(\lambda) =   \mathbf{X}^{(0)}_{B,\pm\omega} + \lambda  \mathbf{X}^{(1)}_{B,\pm\omega} + \lambda^2  \mathbf{X}^{(2)}_{B,\pm\omega} + ...
\end{equation}
From Eq. \eqref{eq:response_function}, the response function corrections up to second order in $\lambda$ thus read
\begin{eqnarray}
     \llangle A ; B \rrangle _\omega ^{(0)} &=& {\tilde{\mathbf{A}}^{(0)\dagger}} \mathbf{X}^{(0)}_{B,\pm\omega} \\ 
     \llangle A ; B \rrangle _\omega ^{(1)} &=& {\tilde{\mathbf{A}}^{(1)\dagger}} \mathbf{X}^{(0)}_{B,\pm\omega} + {\tilde{\mathbf{A}}^{(0)\dagger}} \mathbf{X}^{(1)}_{B,\pm\omega} \\
      \llangle A ; B \rrangle _\omega ^{(2)} &=& {\tilde{\mathbf{A}}^{(2)\dagger}} \mathbf{X}^{(0)}_{B,\pm\omega} + {\tilde{\mathbf{A}}^{(1)\dagger}} \mathbf{X}^{(1)}_{B,\pm\omega} + {\tilde{\mathbf{A}}^{(0)\dagger}} \mathbf{X}^{(2)}_{B,\pm\omega}
\end{eqnarray}
As written, the response function corrections through second order so far require corrections to the property and response vectors also through second order.
Following Eq. \eqref{eq:property_vector}, the corrections to the property vector are 
\begin{eqnarray}
    \tilde{\mathbf{B}}^{(0)} &=& \mathbf{B} \mathbf{c}^{(0)} - \mathbf{c}^{(0)} B^{(0,0)}  \\
   \tilde{\mathbf{B}}^{(1)} &=& \mathbf{B} \mathbf{c}^{(1)} - \mathbf{c}^{(0)} \left( B^{(1,0)} +  B^{(0,1)}\right) - \mathbf{c}^{(1)} B^{(0,0)} \\
   \tilde{\mathbf{B}}^{(2)} &=&  \mathbf{B} \mathbf{c}^{(2)} - \mathbf{c}^{(0)} \left( B^{(2,0)} +  B^{(1,1)} +  B^{(0,2)}\right) - \mathbf{c}^{(1)} \left(B^{(1,0)}+B^{(0,1)}\right)  - \mathbf{c}^{(2)} B^{(0,0)}  
\end{eqnarray}
Where $B^{(n,m)} = \mathbf{c}^{(n)\dagger} \mathbf{B} \mathbf{c}^{(m)} $ is a moment between the wave function corrections of order $n$ and $m$.
We now proceed to express the $\lambda$-dependent linear response equation (see Eq. \eqref{eq:lr_equation}) as
\begin{equation}
\left( \mathbf{H}_0 + \lambda \mathbf{V} - (E_0(\lambda) \pm \omega)\mathbf{I} \right) \mathbf{X}_B(\pm \omega)(\lambda) = \tilde{\mathbf{B}}(\lambda).
\end{equation}
By expanding all the quantities in $\lambda$ and collecting orders, we arrive at a set of linear equations for the $n$'th order correction to the response vectors
\begin{eqnarray}
\left( \mathbf{H}_0 - (E_0^{(0)} \pm \omega)\mathbf{I} \right) \mathbf{X}^{(0)}_B(\pm \omega) &=& \tilde{\mathbf{B}}^{(0)} \label{eq:lr_pert_0} \\
\left( \mathbf{H}_0 - (E_0^{(0)} \pm \omega)\mathbf{I} \right) \mathbf{X}^{(1)}_B(\pm \omega) &=& \tilde{\mathbf{B}}^{(1)} - \mathbf{V}\mathbf{X}^{(0)}_B(\pm \omega)  \\
\left( \mathbf{H}_0 - (E_0^{(0)} \pm \omega)\mathbf{I} \right) \mathbf{X}^{(2)}_B(\pm \omega) &=& \tilde{\mathbf{B}}^{(2)} - \mathbf{V}\mathbf{X}^{(1)}_B(\pm \omega) + E_0^{(2)} \mathbf{X}^{(0)}_B(\pm \omega)  \label{eq:lr_pert_2}
\end{eqnarray}
where we have used the fact that the first-order correction to the energy $E_0^{(1)}$ is zero with an Epstein-Nesbet partitioning.
As was the case for the wave function corrections, Eqs. \eqref{eq:lr_pert_0}-\eqref{eq:lr_pert_2} should be understood to have two sectors. In the internal one, a linear equation is solved using $\mathbf{H}_0$ using an iterative solver routine.
In the external space, $\mathbf{H}_0$ is diagonal, and the response vector coefficients can be evaluated algebraically (similar to how $\mathbf{c}^{(1)}$ is evaluated).

At this point, we must note a slightly unfortunate consequence of the developed theory, namely that the pole structure at finite order remains unchanged relative to the parent variational LR-SCI theory. To see this, note that poles in Eqs. \eqref{eq:lr_pert_0}-\eqref{eq:lr_pert_2} only occur when $\det\left( \mathbf{H}_0 - (E_0^{(0)} \pm \omega)\mathbf{I} \right)$ vanishes. However, this clearly coincides with the standard LR-SCI. As a result, the perturbative treatment does not improve the description of frequency-dependent properties. Although this is unfortunate, the method may yet find use for static linear response properties.
Looking forward, it is possible that alternative theoretical paths, based on a Lagrangian quasi-energy formulation\cite{christiansen1998response} or other alternative theoretical frameworks, might not share this drawback of the currently considered method.  
Exploring such frameworks would be an interesting and important direction for future work, in particular when targeting frequency-dependent properties.

As written, the response second-order correction to the response function requires knowledge of both the second order correction to the property gradient ($\tilde{\mathbf{A}}^{(2)}$) and the second order correction to the response vector ($\mathbf{X}^{(2)}_{B,\pm \omega}$). 
However, by projecting on the left with $\mathbf{X}_A^{(0)\dagger}$ on the second order equation (Eq. \eqref{eq:lr_pert_2}):
\begin{equation}
\mathbf{X}_A^{(0)\dagger} \left( \mathbf{H}_0 - (E_0^{(0)} \pm \omega)\mathbf{I} \right) \mathbf{X}^{(2)}_B(\pm \omega) = \mathbf{X}_A^{(0)\dagger}  \left( \tilde{\mathbf{B}}^{(2)} - \mathbf{V}\mathbf{X}^{(1)}_B(\pm \omega) + E_0^{(2)} \mathbf{X}^{(0)}_B(\pm \omega)  \right) 
\end{equation}
and noting that $ \left( \mathbf{H}_0 - (E_0^{(0)} \pm \omega)\mathbf{I} \right)$ is symmetric, we may let it act to the left, which by Eq. \eqref{eq:lr_pert_0} gives
\begin{equation}
\tilde{\mathbf{A}}^{(0)\dagger}  \mathbf{X}^{(2)}_B(\pm \omega) = \mathbf{X}_A^{(0)\dagger}  \left( \tilde{\mathbf{B}}^{(2)} - \mathbf{V}\mathbf{X}^{(1)}_B(\pm \omega) + E_0^{(2)} \mathbf{X}^{(0)}_B(\pm \omega)  \right) 
\end{equation}
so we can rewrite
\begin{equation}
          \llangle A ; B \rrangle _\omega ^{(2)} = {\tilde{\mathbf{A}}^{(2)\dagger}} \mathbf{X}^{(0)}_{B,\pm\omega} + {\tilde{\mathbf{A}}^{(1)\dagger}} \mathbf{X}^{(1)}_{B,\pm\omega} +  \mathbf{X}_A^{(0)\dagger}  \left( \tilde{\mathbf{B}}^{(2)} - \mathbf{V}\mathbf{X}^{(1)}_B(\pm \omega) + E_0^{(2)} \mathbf{X}^{(0)}_B(\pm \omega)  \right) 
\end{equation}
Thus, the second-order correction to the response vector is not explicitly needed, at the expense of storing the products $\mathbf{V}\mathbf{X}_B^{(1)}$.
In this formulation, however, the second-order correction to the property vector is still required, which in turn requires the second-order correction to the wave function coefficients, $\mathbf{c}^{(2)}$.

\section{Computational Details}
The LR-SCI implementation described in this work is available from \url{https://github.com/peter-reinholdt/sci-resp/tree/common-space}.
We used the PyCI library\cite{richer2024pyci} for all HCI-related routines, including determinant selection, sparse Hamiltonian construction, and Hamiltonian matrix-vector multiplication.
We added local modifications (available at \url{https://github.com/peter-reinholdt/PyCI/tree/V-matvec}) to enable the calculation of direct $\hat{V}$ operator matrix-vector product routines necessary for the perturbative treatment.
All variational LR-SCI calculations have been conducted using the GS+V+X scheme with common screening thresholds  $\varepsilon_1 \equiv \varepsilon_0 = \varepsilon_V = \varepsilon_X$ for the ground-state, property vector, and response vector determinant addition steps.
To validate the correctness of the implementation, we performed finite-difference calculations on small test systems using the $\lambda$-dependent Hamiltonian. Quantities such as the energies, the wave function, property vectors, and response vectors were evaluated at finite $\lambda$ on a grid of equally spaced $\lambda$-values. The perturbatively corrected quantities were extracted as numerical derivatives using a 7-point stencil and were compared with the corresponding analytical perturbative expressions order by-order. 

Molecular geometries of water, BH, and HCl were taken from the QUEST database\cite{loos2025quest}.
The structure of ethene was optimized with frozen-core CCSD(T)/aug-cc-pVTZ using Orca, version 6.0.0\cite{neese2025software}.
Reference FCI calculations were obtained using the Dalton program\cite{daltonpaper}, while CCSD, CCSDT, and CCSDTQ calculations were carried out with CFOUR\cite{matthews2020coupled} and MRCC\cite{mester2025overview}.
Hartree-Fock calculations and integrals were obtained using PySCF\cite{pyscf}.

\section{Results and Discussion}
\subsection{The Polarizability of Water}
\begin{figure}
    \centering
    \includegraphics[width=.8\linewidth]{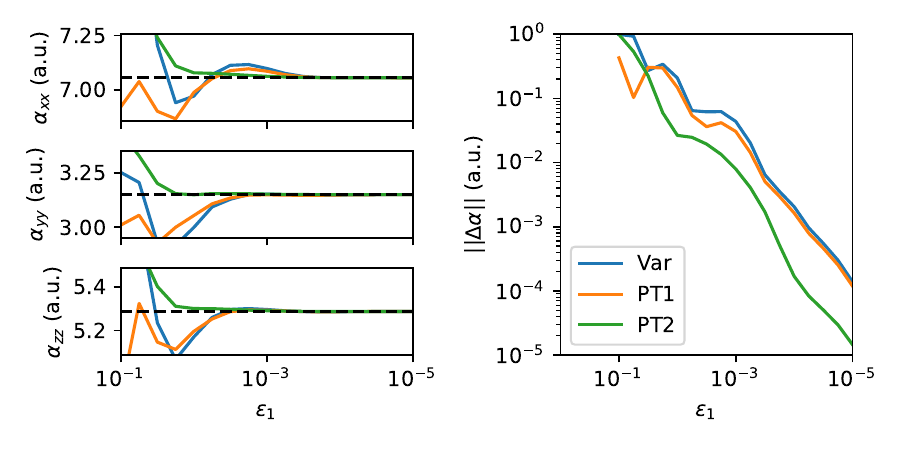}
    \caption{Convergence of the LR-SCI-PT polarizability of water/cc-pVDZ (8e, 23o) towards the CASCI limit. The left panels show the polarizability components on a linear scale, while the right panel shows the error in the polarizability on a logarithmic scale. The parameter $\varepsilon_1$ controls the accuracy of the zeroth-order LR-SCI calculation. }
    \label{fig:water_pt2_convergence}
\end{figure}

We begin by assessing the performance of the LR-SCI-PT approach for a system where exact reference data (FCI) is accessible, allowing for a direct evaluation of the accuracy of the perturbative corrections.
Figure~\ref{fig:water_pt2_convergence} shows the convergence of the components of the LR-SCI-PT polarizability for the water molecule (cc-pVDZ basis, frozen core). The results are shown as a function of the parameter $\varepsilon_1$, which controls the accuracy of the zeroth-order LR-SCI calculation (GS+V+X). Smaller values of $\varepsilon_1$ are more accurate, but include more determinants, increasing the computational cost.
We track the convergence of the variational result, the first-order-corrected result (PT1), and the second-order-corrected result (PT2).
The variational LR-SCI and the perturbatively corrected calculations converge systematically towards the complete active space configuration interaction (CASCI) limit as $\varepsilon_1$ is tightened. 
The first-order correction (PT1, orange lines) to the polarizability is rather small, yielding only very small improvements in accuracy over the zeroth-order calculation. In contrast, including second-order corrections (PT2, green lines) substantially improves accuracy, typically by one order of magnitude (i.e., one extra digit of accuracy).
It is also interesting to note that the oscillatory convergence behaviour in the zeroth-order method (converging from above, then from below, then from above, etc.) is diminished for the PT2-corrected polarizability.
Targeting an accuracy of approximately $10^{-3}$ a.u. relative to the FCI reference can be achieved with the variational approach using $\varepsilon_1 = 10^{-4.25}$ ($2.4\times10^{6}$ determinants), whereas the perturbative approach manages a similar accuracy with a looser threshold $\varepsilon_1 = 10^{-3.50}$ ($0.43\times10^{6}$ variational determinants). 

\subsection{Approximations to the Perturbative Treatment}
So far, we have used an exact treatment of perturbative corrections. For practical calculations, it will be more computationally efficient to introduce approximations to the size of the perturbative space and of $\hat{V}$ matrix-vector products.
We introduce a parameter $\varepsilon_2$ that controls the accuracy of the perturbative description, similar to how perturbative treatments are considered in HCI\cite{holmes2016heat}. We only include perturbative determinants when $\left|V_{IJ} c_J\right| > \varepsilon_2$, 
and the $\hat{V}$ matrix-vector products are approximated as
\begin{equation}
\sum_J V_{IJ} c_J \approx \sum_{\substack{J \\ |V_{IJ} c_J| > \varepsilon_2}} V_{IJ} c_J,
\end{equation}
i.e., the matrix-vector products only include terms above a certain magnitude.
We consider two schemes for the perturbative accuracy parameter $\varepsilon_2$.
In the first scheme, we let $\varepsilon_2$ control both the inclusion of perturbative determinants and the accuracy of the $\hat{V}$ matrix-vector products.
In the second scheme, we use two different thresholds, one $\varepsilon_2$ controlling which perturbative determinants to keep track of, and another $\varepsilon_2^{\mathrm{mult}}$ which controls the accuracy of the $\hat{V}$ matrix-vector multiplication. The motivation for this second scheme is mainly concerned with achieving a good balance between the memory footprint and accuracy of the method: if we decide to commit memory to storing information about a perturbative determinant, it may be worthwhile to spend slightly more computational power to ensure an accurate representation of the matrix-vector products.  

\begin{figure}
    \centering
    \includegraphics[width=1.0\linewidth]{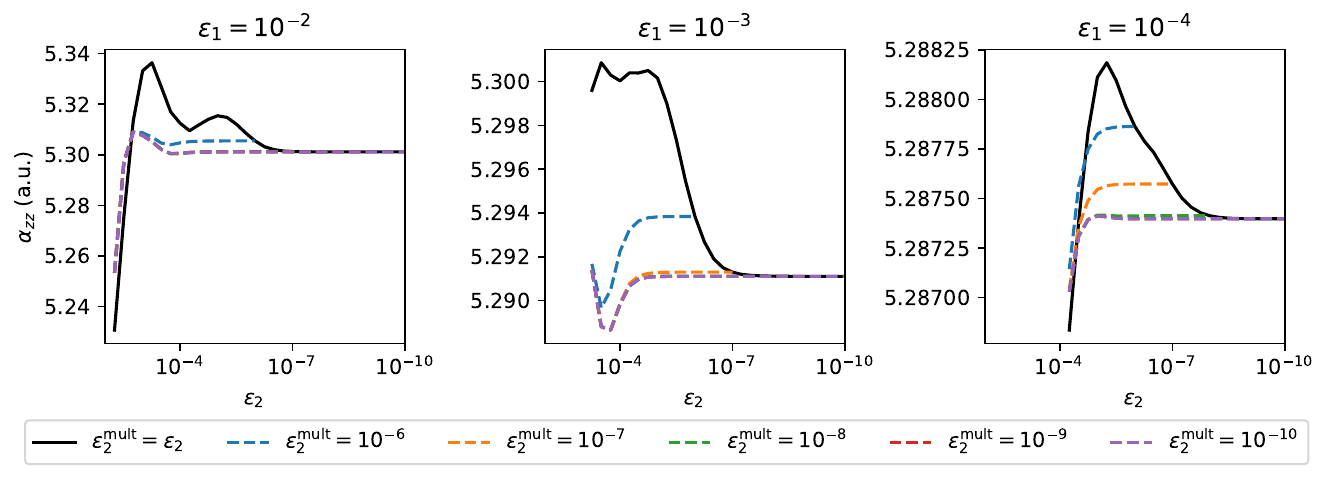}
    \caption{Convergence of the $zz$ component of the ground-state polarizability of water/cc-pVDZ (8e, 23o) as a function of the perturbative accuracy parameter $\varepsilon_2$. The three panels show results for different choices of the variational space parameter $\varepsilon_1$. The black lines show a version in which the parameter $\varepsilon_2$ controls both the accuracy of the fluctuation operator matrix-vector multiplication and the size of the perturbative space. The colored dashed lines show a version in which $\varepsilon_2$ controls only the size of the perturbative space, while the accuracy of the fluctuation operator matrix-vector multiplication is manually controlled (see legend).}
    \label{fig:water_eps2_convergence}
\end{figure}

We consider the effect of these approximations in Figure \ref{fig:water_eps2_convergence}, which shows the convergence of $\alpha_{zz}$ of water with respect to the perturbative accuracy parameter $\varepsilon_2$.
The results are given for three choices of the variational space parameter $\varepsilon_1$.
We note that the scales of the three panels differ: as the reference space improves (smaller $\varepsilon_1$), the magnitude of the perturbative correction decreases, and thus the overall variation with respect to $\varepsilon_2$ becomes correspondingly smaller.
We first consider the scheme in which the $\hat{V}$ matrix-vector multiplication and perturbative space sizes are both controlled by $\varepsilon_2$ (black lines). In this case, we find that good convergence of the perturbative correction requires $\varepsilon_2 \approx 10^{-7}$ for $\varepsilon_1=10^{-2}$ and  $\varepsilon_1=10^{-3}$, while a tighter $\varepsilon_2\approx 10^{-8}$ would be required for the $\varepsilon_1=10^{-4}$ calculation.
Overall, these are quite stringent requirements, and the convergence of the perturbative correction with respect to $\varepsilon_2$ is rather slow in the first scheme.
We next consider the second scheme, in which the determinant-selection and matrix-vector multiplication thresholds are decoupled, as shown with the colored lines in Figure \ref{fig:water_eps2_convergence}.
We find that good convergence of the perturbative correction can indeed be obtained with such a hybrid scheme. For example, for $\varepsilon_1=10^{-2}$, fixing the multiplication routine to an accuracy of $\varepsilon_2^{\mathrm{mult}}=10^{-7}$ allows using a perturbative space of just $\varepsilon_2 = 10^{-4}$. 
This suggests that the somewhat slow convergence in $\varepsilon_2$ is mainly driven by inaccurate perturbative matrix-vector products, and not from the neglect of many weakly contributing determinants.
For the two remaining cases, we find that good convergence of the perturbative correction is obtained with a combination of $\varepsilon_1,\varepsilon_2,\varepsilon_2^{\mathrm{mult}}$ of $10^{-3},10^{-5},10^{-8}$ and  $10^{-4},10^{-6},10^{-9}$, which overall suggests that a balanced description is obtained with $\varepsilon_2 = 10^{-2}\cdot \varepsilon_1$ and $\varepsilon_2^{\mathrm{mult}} = 10^{-5}\cdot \varepsilon_1$. We have adopted these thresholds for the following calculations.

For further computational efficiency improvements, a promising alternative to the deterministic approach described above would be to adopt a semi-stochastic framework similar to the approach used in semistochastic HCI\cite{sharma2017semistochastic,li2018fast}, although whether such an approach would be applicable to the more complicated structure of the linear response equations is yet to be determined. 

\subsection{The Polarizability of Ethene}

\begin{figure}
    \centering
    \includegraphics[width=0.4\linewidth]{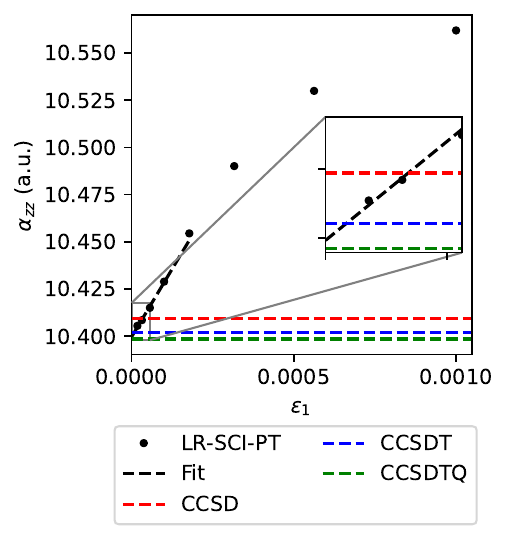}
    \caption{The $\alpha_{zz}$ component of the polarizability of ethene/cc-pVDZ (12e, 46o). The polarizability is computed using LR-SCI-PT for a set of increasingly tight thresholds $\varepsilon_1$. The inset shows a zoomed region around $\varepsilon_1=0$. The black dashed line shows an extrapolation towards $\varepsilon_1=0$. Results obtained using CCSD, CCSDT, and CCSDTQ are indicated with dashed lines.}
    \label{fig:ethene_alpha_zz}
\end{figure}

We next consider the ethene molecule in a cc-pVDZ basis (12e, 46o), which corresponds to an FCI dimension of $8.77 \times 10^{13}$.
We have oriented the ethene molecule with the molecular plane in the $xy$ plane, and the carbon-carbon bond along the $x$-axis, and $z$ orthogonal to the plane of the molecule.
Figure \ref{fig:ethene_alpha_zz} shows the $\alpha_{zz}$ component of the polarizability of ethene, computed using the perturbatively corrected LR-SCI for a series of $\varepsilon_1$ values, plotted on a linear scale.
For comparison purposes, results obtained with CCSD (red), CCSDT (blue), and CCSDTQ (green) are indicated by dashed lines.
At loose $\varepsilon_1$ ($10^{-3}$), the LR-SCI-PT polarizability overestimates the polarizability by about 0.16 a.u. 
As $\varepsilon_1$ is lowered, the accuracy of the computed polarizability improves, and it eventually agrees quite well with the results obtained using coupled-cluster methods. 
For the final 3--4 points, the polarizability appears to depend nearly linearly on $\varepsilon_1$, and we thus take the liberty of doing a linear fit and extrapolating towards the $\varepsilon_1=0$ limit, which gives an estimate of 10.3996 a.u. for $\alpha_{zz}$, which is in quite good agreement with CCSDTQ (10.3984 a.u.).
This extrapolated LR-SCI-PT value is slightly lower than the results obtained from CCSD (10.409 a.u.) and CCSDT (10.402 a.u.).
The raw, un-extrapolated value (10.405) obtained with LR-SCI-PT at $\varepsilon_1=10^{-4.75}$ ``only'' manages accuracy comparable to CCSD.
The largest LR-SCI-PT calculation considered contained $65$ million variational determinants and $11 \times10^{9}$ perturbative determinants, and took 27 hours on a compute node equipped with 2x AMD 7742 64-Core CPUs (128 cores in total). 
The underlying variational LR-SCI calculation consumes about 20\% of the computational time (about 5 hours).
For the perturbative part, the calculation is dominated by the calculation of the fluctuation operator matrix-vector products, which take up around 70\% of the time in the perturbative routines.
For comparison, the CCSDTQ calculation took about 18 hours using 16 cores, and we may thus note that for weakly correlated systems, coupled-cluster-based methods are computationally more efficient than the present LR-SCI-PT implementation.

\subsection{Ground- and Excited State Polarizabilities}
We now turn to consider the calculation of excited-state polarizabilities. Within the LR-SCI framework, excited-state response properties are naturally accessible, as any eigenstate can serve as the reference in the linear-response equations (and the following perturbative treatment).
This is in contrast to traditional response theory starting from the electronic ground state, where evaluation of the excited-state polarizability requires evaluating a double residue of a cubic response function\cite{helgaker2012recent}. 
\citeauthor{naim2026excited} recently reported benchmark calculations of ground- and excited-state polarizabilities using CC3 as a reference method\cite{naim2026excited}, and used those calculations to evaluate the performance of several lower-accuracy wave-function as well as density-functional-based methods.
Here, we repeat selected calculations from that benchmark study.
For the excited states, there is occasionally some ambiguity about which root is targeted, since different eigenstates may change ordering depending on which determinants are included in the CI expansion. We address this problem by first converging to a desired excited state with $\varepsilon_0=10^{-4}$, with assignment of the state made by inspection of the CI vector. After the desired root is identified, we then start all the following LR-SCI calculations from this initial wave function. When the determinant space is expanded in later LR-SCI steps, we keep track of the desired root by computing the overlap with the initial reference calculation.

We first consider the BH molecule, which is a small enough system that an exact FCI treatment is feasible.
\begin{figure}
    \centering
    \includegraphics[width=.8\linewidth]{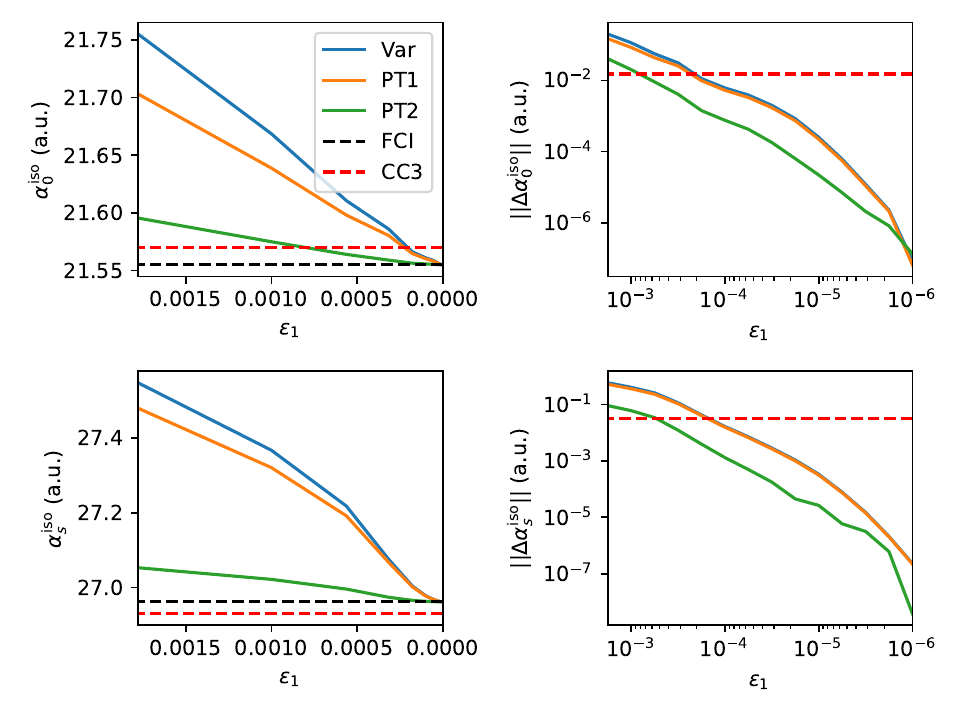}
    \caption{Convergence of LR-SCI methods towards the FCI limit for the isotropic ground- ($X~^1\Sigma^+$, top panels) and excited-state ($1~ ^1\Pi$, bottom panels) polarizabilities BH/aug-cc-pVTZ (4e, 68o). The left panels plot the polarizabilities as a function of the accuracy parameter $\varepsilon_1$ on a linear scale. The right panels show the errors (relative to FCI) on a logarithmic scale. The values for the CC3 polarizability were taken from Ref. \citenum{naim2026excited}.}
    \label{fig:BH_es_convergence}
\end{figure}
As shown in Figure \ref{fig:BH_es_convergence}, we find that both variational and perturbatively corrected LR-SCI converge systematically to the FCI limit as $\varepsilon_1$ is tightened. 
Similar to the previously discussed calculations on the polarizability of water, we find that going from the variational LR-SCI theory to the first-order perturbatively corrected one (PT1) leads only to relatively minor improvements in accuracy, whereas including second-order corrections gives a more substantial increase in accuracy.
The CC3 estimate of the polarizability of BH is quite accurate, with the ground-state polarizability being underestimated by about 0.02 a.u., while the excited-state polarizability is overestimated by about 0.03 a.u.
Aiming for an accuracy of 0.01 a.u. in the polarizability, LR-SCI-PT requires $\varepsilon_1$ of about $3\times 10^{-4}$ with LR-SCI-PT (or $7\times10^{-5}$ with variational LR-SCI), while a higher accuracy of 0.001 a.u. requires an $\varepsilon_1$ of $8\times10^{-5}$ (or $1.7\times 10^{-5}$ with variational LR-SCI).
Overall, the BH results confirm that excited-state polarizabilities can be treated on the same footing as ground-state properties within the LR-SCI-PT framework, with systematic convergence towards the FCI limit.

\begin{figure}
    \centering
    \includegraphics[width=.8\linewidth]{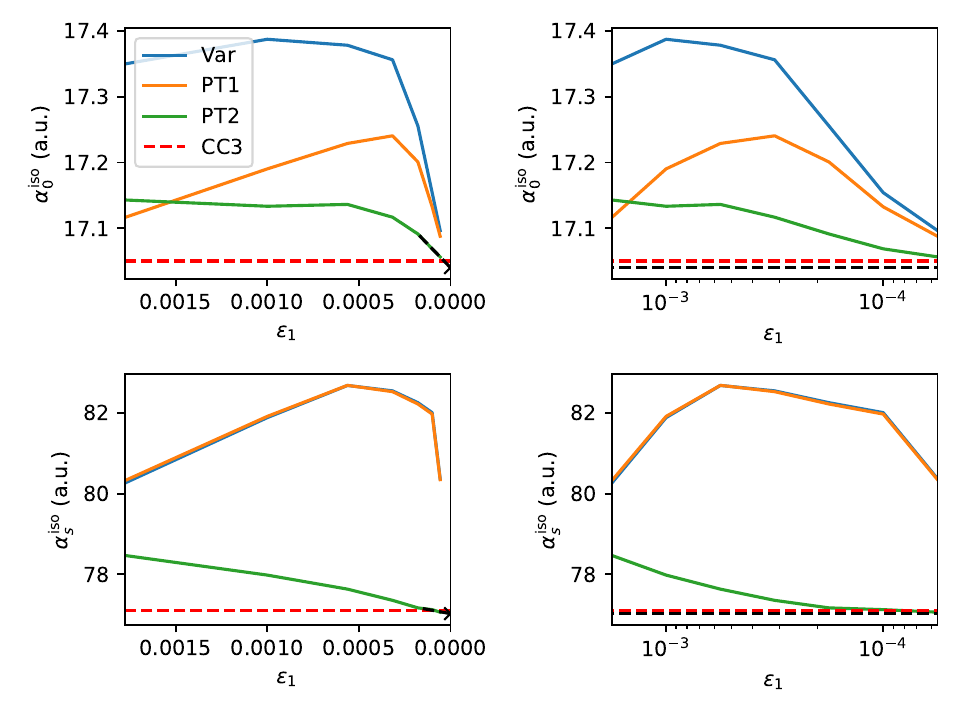}
        \caption{Convergence of LR-SCI methods towards the FCI limit for the isotropic ground- ($X~^1\Sigma^+$, top panels) and excited-state ($1~ ^1\Pi$, bottom panels) polarizabilities HCl/aug-cc-pVTZ (8e, 68o). The left panels plot the polarizabilities as a function of the accuracy parameter $\varepsilon_1$ on a linear scale, while the right panels show the same data on a log-linear scale. The values for the CC3 polarizability were taken from Ref. \citenum{naim2026excited}. The black dotted line indicates the result of a linear extrapolation towards $\varepsilon_1=0$. }
    \label{fig:HCl_es_convergence}
\end{figure}

Next, we consider the hydrogen chloride molecule with the aug-cc-pVTZ (8e, 68o), as shown in Figure \ref{fig:HCl_es_convergence}.
The largest LR-SCI calculation used $\varepsilon_1=10^{-4.25}$, which corresponds to 18.7 million variational with 2150 million perturbative determinants for the ground state, and 97.0 million variational with 6700 million perturbative determinants for the excited states.
The polarizability of the ground state is predicted to be 17.056 a.u. with the second-order perturbatively corrected LR-SCI calculation, and an extrapolation of the last three points towards $\varepsilon_1 = 0$ gives an estimate of 17.040 a.u. Both of these estimates appear in reasonable agreement with the CC3 polarizability (17.05 a.u.).
For the excited state, we obtain a polarizability of 77.061 a.u. with the second-order perturbatively corrected LR-SCI method, while the extrapolated value is 77.024 a.u. Again, both of these are in quite reasonable agreement with the CC3 polarizability (77.10 a.u.). 
For the excited state, the raw variational LR-SCI result (80.371 a.u.) is notably quite far from both the second-order perturbatively corrected LR-SCI and the extrapolated result, as well as from the CC3 polarizability. 
Overall, this demonstrates that the perturbative correction substantially enhances the accuracy of LR-SCI, enabling reliable predictions even for systems beyond the reach of the fully variational treatment.
Our results also suggest that the CC3 method provides highly accurate estimates of excited-state polarizabilities for such systems.

\subsection{Dissociation Curves}
So far, we have considered the convergence characteristics of LR-SCI-PT for mostly weakly correlated systems around equilibrium geometries.
We next consider a more demanding test with the description of molecular properties along bond-dissociation coordinates. 
Bond breaking can present challenges for electronic-structure methods\cite{sherrill2005bond}, since stretched geometries and the breaking/formation of chemical bonds typically feature near-degenerate electron configurations, more pronounced static correlation, or multiconfigurational character.

\begin{figure}
    \centering
    \includegraphics[width=0.5\linewidth]{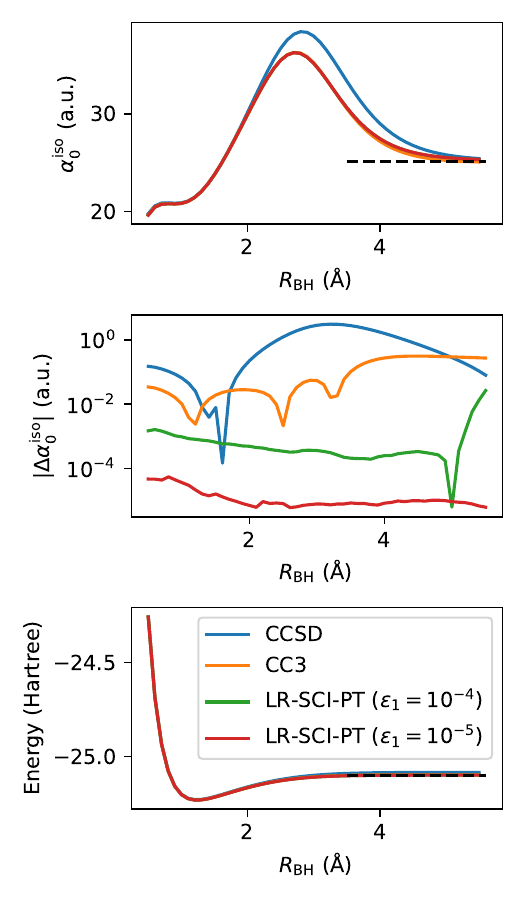}
    \caption{Dissociation curves of BH/aug-cc-pVTZ (4e, 68o) computed with CCSD, CC3, and LR-SCI-PT. The top panel shows the isotropic polarizability, while the middle panel shows the absolute error in the polarizability (relative to the CASCI reference) on a logarithmic scale. The bottom panel shows the potential energy curve. The black dashed lines indicate the dissociation limit, calculated as the sum of the energies or polarizabilities of the isolated B and H atoms.}
    \label{fig:BH-curve}
\end{figure}

Figure \ref{fig:BH-curve} shows the isotropic polarizability of the BH molecule as a function of the B--H bond distance, computed using either coupled-cluster methods or LR-SCI-PT.
Around the potential-energy minimum (1.2 \AA), the isotropic polarizability is around 21.6 a.u. As the B--H bond distance increases, the polarizability increases steadily until a distance of 2.7 \AA, where it reaches a maximum of 36.3 a.u. The polarizability decreases with further extension of the B--H bond, eventually approaching the isolated-atom limit (black dashed lines).
On the plotted scale, LR-SCI-PT with $\varepsilon_1$ of $10^{-4}$ to $10^{-5}$ appears virtually identical and are indistiguishable from the CASCI reference.
Both CCSD and CC3 also give a quite accurate reproduction of the polarizability for this system, especially around the equilibrium bond distance. However, at the distance of 2.7 \AA, near the maximum polarizability, CCSD overestimates the polarizability by 1.9 a.u., and just before reaching the dissociation limit, CC3 very slightly underestimates the polarizability.
The small errors are more clearly seen on the logarithmic scale (Figure \ref{fig:BH-curve}, middle panel), where we find that LR-SCI-PT with $\varepsilon_1=10^{-4}$ gives errors below 0.002 a.u., except near the dissociation limit where they increase to 0.02 a.u., while the tighter threshold of  $\varepsilon_1=10^{-5}$ gives errors below $10^{-4}$ a.u. throughout the dissociation curve.
From the potential-energy curve (Figure \ref{fig:BH-curve}, bottom panel), it is clear that the dissociation of the BH molecule is qualitatively correct for all the considered methods, including CCSD, and that the dissociation of this single bond is non-problematic for the considered methods. 

\begin{figure}
    \centering
    \includegraphics[width=0.5\linewidth]{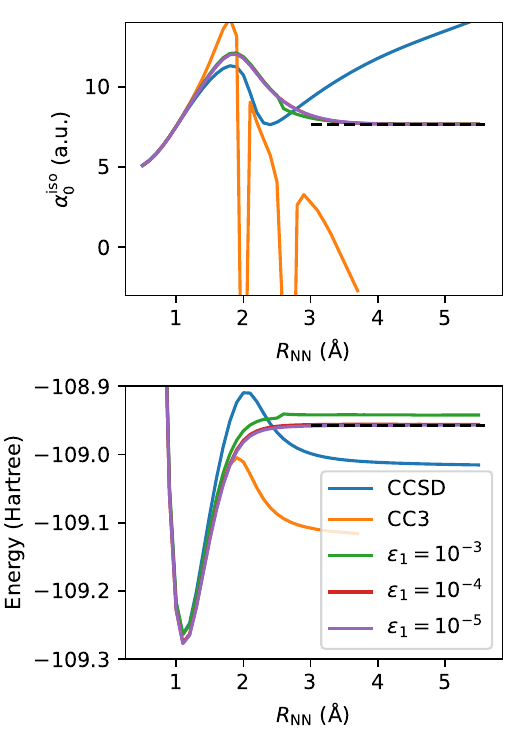}
    \caption{Dissociation curves of N$_2$/cc-pVDZ (10e, 26o) computed with CCSD, CC3, and LR-SCI-PT. The top panel shows the isotropic polarizability, while the bottom panel shows the potential energy curve. The black dashed lines indicate the dissociation limit, calculated as the sum of the energies or polarizabilities of the isolated N atoms.}
    \label{fig:N2-curve}
\end{figure}

Next, we consider a more difficult example, with the dissociation of the triple bond in N$_2$, as shown in Figure \ref{fig:N2-curve}.
The isotropic polarizability of N$_2$ follows a course qualitatively similar to that of the BH molecule: around the equilibrium distance (1.1 \AA), the polarizability is relatively small (around 8.2 a.u.), and increases until a maximum value (12.0 a.u.) at a distance of 1.9 \AA, after which further extension of the N--N bond decreases the polarizability until it eventually approaches the dissociation limit, where the polarizability becomes equal to the sum of the isolated-atom polarizabilities (black dashed lines), 7.7 a.u.
This behaviour is captured quite well by LR-SCI-PT methods, where even the relatively loose criterion of $\varepsilon=10^{-3}$ gives a mostly qualitatively correct description, except near 2.6 \AA, where the polarizability displays an unphysical kink. A similar kink is also visible in the associated potential energy curve (Figure \ref{fig:N2-curve}, bottom panel).
The unphysical kink is due to the discrete nature of including determinants in the SCI wave function, i.e., that a determinant is either included in the CI expansion or it is not. Inspecting the energy curves of the underlying SCF and the PT2 energy correction, both curves appear smooth; however, the variational SCI calculation is not.
A highly accurate description requires a tighter threshold of at least  $\varepsilon_1=10^{-4}$.

The considered coupled-cluster methods give a good description of the polarizability near the equilibrium distance, but give a physically incorrect behaviour when the N--N bond distance is increased beyond 1.5 \AA. For CCSD, the polarizability is underestimated by 0.8 a.u. at a distance of 1.9 \AA, and has a minimum at 2.4 \AA, after which the polarizability starts to increase again toward a value larger than the sum of the isolated polarizabilities of the N atoms. 
For CC3, extension beyond 1.8 \AA~leads to a somewhat erratic description of the polarizability, occasionally even becoming negative.
The potential energy curve (Figure \ref{fig:N2-curve}, bottom panel) also shows a corresponding unphysical behaviour for the coupled-cluster methods, with a ``barrier'' at 1.9 \AA~(CCSD) or 2.0~\AA (CC3), followed by the energy going below the variational limit at larger bond extensions.
Additionally, we note that solving the coupled-cluster ground-state equations became difficult at longer bond distances.
Initially, the only consequence is simply an increase in the number of iterations required to converge, while at large bond extensions, especially with CC3, we were ultimately unable to obtain convergence at distances beyond 3.7 \AA. 
Thus, breaking the N-N triple bond is evidently challenging, and the coupled-cluster methods considered are unable to provide a physically reasonable description of systems with strong static correlation.

Overall, these results show that bond breaking is not universally difficult for coupled-cluster methods, as illustrated by the accurate description of BH dissociation. 
However, more challenging cases, such as breaking the triple bond of N$_2$, where substantial static correlation develops upon bond extension, can lead to qualitative failures in the considered coupled-cluster approximations. LR-SCI-PT, in contrast, retains a systematically improvable description in both regimes.

\FloatBarrier
\section{Conclusion}\label{sec:conclusion}
In this work, we have formulated a scheme for perturbative corrections to the LR-SCI method, which corrects molecular response properties for missing correlation effects due to a limited-size variational determinant space using an Epstein–Nesbet perturbation expansion through second order.
The method is based on an order-by-order perturbation expansion of the linear response function.
One of the central computational costs present in the method is the computation of fluctuation operator matrix-vector products, and we introduce and discuss approximate yet accurate schemes to evaluate these matrix-vector products.
A central and slightly disappointing theoretical finding is that the pole structure at finite order remains identical to the parent variational LR-SCI theory, which means that the method in practice is useful only for static response properties and that it is unsuitable for the treatment of frequency-dependent properties.

We carry out numerical tests targeting static polarizabilities of water, ethene, boron hydride, and hydrogen chloride to validate the precision and convergence characteristics of the method.
For small systems, where comparisons to FCI remain feasible, we find systematic convergence towards the FCI limit for both LR-SCI and the perturbatively corrected variants. We show that the first-order perturbative correction typically provides only marginal improvements, and that inclusion of second-order perturbative corrections is required to obtain enhanced accuracy over the underlying variational calculation. Further, we find that the occasional oscillatory convergence behaviour present in the reference variational calculations appears to be diminished upon the inclusion of second-order perturbative corrections.
For larger molecular systems such as ethene, a linear extrapolation of the perturbatively corrected polarizabilities toward the $\varepsilon_1 \to 0$ limit yields excellent agreement with high-level coupled-cluster methods.

Finally, we apply the perturbatively corrected LR-SCI method to calculate ground- and excited-state isotropic polarizabilities of $\text{BH}$ and $\text{HCl}$, and show that the second-order perturbative correction successfully brings tight convergence even when the raw variational reference is notably far from the limit. We also consider the calculation of polarizabilities along bond-dissociation coordinates, using the BH and N$_2$ molecules as examples, and show that for systems featuring substantial static correlation, LR-SCI-PT is able to provide accurate and physically reasonable results in regimes where coupled-cluster methods show qualitative failures. 
Overall, the $\text{LR-SCI-PT}$ methodology provides a systematic and powerful route toward near-$\text{FCI}$ quality molecular properties for chemical systems beyond the reach of exact FCI treatments.

\begin{acknowledgement}
Computations/simulations for the work described herein were supported by the DeIC National HPC Centre, SDU.
\end{acknowledgement}


\section*{Data and Software Availability Statement}
The LR-SCI-PT implementation described in this work is available from \url{https://github.com/peter-reinholdt/sci-resp/tree/common-space}. It relies on a locally modified version of the PyCI library\cite{richer2024pyci} available at \url{https://github.com/peter-reinholdt/PyCI/tree/V-matvec}.
Data generated and analyzed in this study are included in the published article.
\bibliography{main}
\end{document}